\documentclass[pre,twocolumn,showpacs,preprintnumbers,amsmath,amssymb]{revtex4}
\usepackage{graphicx}% Include figure files
\usepackage{dcolumn}% Align table columns on decimal point
\usepackage{bm}% bold math

\begin{document}

\author{D.E. Holz$^1$, H. Orland$^{1,2}$, and A. Zee$^{1,3}$ \\
%EndAName
$^{1}$Institute for Theoretical Physics,\\
University of California, Santa Barbara, CA 93106, USA \\
$^{2}$Service de Physique Th\'eorique, CEA-Saclay,\\
91191 Gif-sur-Yvette Cedex, France\\
$^{3}$Center for Advanced Study, Tsinghua University,\\
Beijing 10084, People's Republic of China}
\title{On the Remarkable Spectrum of a Non-Hermitian Random Matrix Model}
%\date{September 26, 2001 }
\begin{abstract}
A non-Hermitian random matrix model proposed a few years ago
has a remarkably intricate spectrum. Various attempts have
been made to understand the spectrum, but even its dimension
is not known. Using the Dyson-Schmidt equation, we show that
the spectrum consists of a non-denumerable set of lines in
the complex plane. Each line is the support of the spectrum
of a periodic Hamiltonian, obtained by the infinite
repetition of any finite sequence of the disorder
variables. Our approach is based on the ``theory of words.''
We make a complete study of all 4-letter words.  The
spectrum is complicated because our matrix contains
everything that will ever be written in the history of the
universe, including this particular paper.
\end{abstract}
\maketitle

\newpage

\section{A Class of Non-Hermitian Random Matrix Models}

Some time ago, Feinberg and one of us (in a paper to be referred to as FZ
\cite{FZ}) proposed the study of the equation
\begin{equation}
\psi _{k+1}+r_{k-1}\psi _{k-1}=E\psi _k  \label{Seq}
\end{equation}
where the real numbers $r_k$ are generated from some random distribution.
Two particularly simple models were studied: (A) the $r_k$'s are equal to $%
\pm 1$ with equal probability, and (B) $r_k=e^{i\theta _k}$ with the angle $%
\theta _k$ uniformly distributed between $0$ and $2\pi .$

Imposing the boundary condition $\psi _{0}=\psi _{N+2}=0$ on (\ref{Seq}) we
can write the set of equations as the eigenvalue equation
\begin{equation}
H_{N}\psi =E\psi
\end{equation}
with $\psi $ the column eigenvector with components $\{\psi _{1},$ $\psi
_{2},$ $\cdots ,$ $\psi _{N},$ $\psi _{N+1}\}$ and $H_{N}$ the $(N+1)$ by $%
(N+1)$ non-Hermitian random matrix
\[
H_{N}=\left(
\begin{array}{llllll}
0 & 1 & 0 &  &  &  \\
r_{1} & 0 & 1 & 0 &  &  \\
0 & r_{2} & 0 & 1 & 0 &  \\
& 0 & \ddots & \ddots & \ddots & 0 \\
&  & 0 & r_{N-1} & 0 & 1 \\
&  &  & 0 & r_{N} & 0
\end{array}
\right)
\]

While quantum mechanics is of course Hermitian it is convenient to think of $%
H$ as a Hamiltonian and (\ref{Seq}) as the non-Hermitian Schr\"{o}dinger
equation describing the propagation of a particle hopping on a 1-dimensional
lattice.

Some applications of non-Hermitian random Hamiltonians include vortex line
pinning in superconductors~\cite{1,2,fz2} and growth models in population
biology \cite{4}. A genuine localization transition can occur
for random non-Hermitian Schr\"{o}dinger Hamiltonians~\cite
{5,6,7,8,9,10,11,12} in one dimension.

As mentioned in FZ, with the open chain boundary condition $\psi _{0}=\psi
_{N+2}=0$ the more general equation
\begin{equation}
s_{k+1}\psi _{k+1}+r_{k-1}\psi _{k-1}=E\psi _{k}  \label{schrody}
\end{equation}
can always be reduced to (\ref{Seq}) by an appropriate ``gauge''
transformation $\psi _{k}\rightarrow \lambda _{k}\psi _{k}.$
Furthermore, applying the transformation $\psi _k\rightarrow u^{-k}\psi _k$
to (\ref{Seq}) we see that if we change $r_k\rightarrow u^2r_k$ then the
spectrum changes by $E\rightarrow uE.$ Thus, scaling the magnitude of the $%
r_k$'s merely stretches the spectrum, and flipping the sign of all the $r_k$%
's corresponds to rotating the spectrum by $90^0.$

It is also useful to formulate the problem in the transfer matrix formalism.
Write (\ref{Seq}) as
\begin{equation}
\left(
\begin{array}{l}
\psi _{k+1} \\
\psi _{k}
\end{array}
\right) =T_{k-1}\left(
\begin{array}{l}
\psi _{k} \\
\psi _{k-1}
\end{array}
\right)
\end{equation}
where the transfer matrix $T_{k}$ is defined as the 2 by 2 matrix
\begin{equation}
T_{k}=\left(
\begin{array}{ll}
E & -r_{k} \\
1 & 0
\end{array}
\right)
\end{equation}
Define $P\equiv T_{N}T_{N-1}\cdots T_{2}T_{1}.$ Then the boundary condition
implies
\begin{equation}
EP_{11}+P_{12}=0
\end{equation}
The solution of this polynomial equation in $E$ determines the spectrum.

\begin{figure}
\resizebox{8cm}{8cm}{
\includegraphics{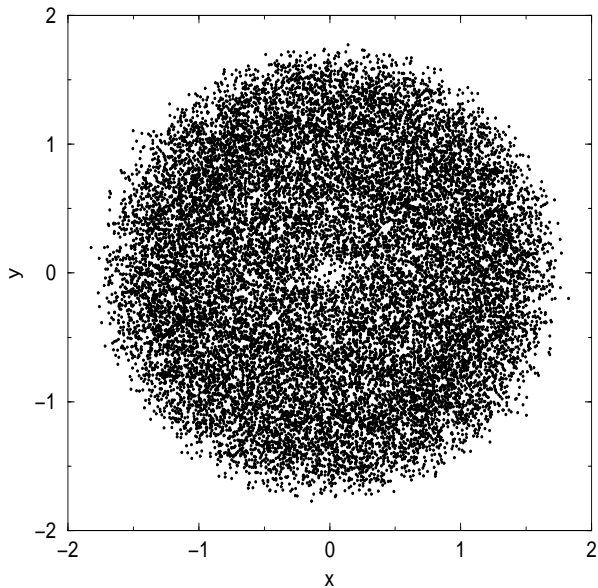}}
\caption{\label{fig1}Support of the density of states for model B.}
\end{figure}

Since $H$ is non-Hermitian the eigenvalues invade the complex plane. For
model B, the spectrum has an obvious rotational symmetry and forms a disk 
(see Fig.~\ref{fig1}, which displays the support of the density of states).
%\begin{figure}[tbh]
%  \epsfxsize=0.5\linewidth
%  \centerline{\hbox{ \epsffile{fig1.eps} }}
%\centerline{Fig.~1: Support of the density of states for model B.}
%\end{figure}
An expansion of the density of eigenvalues around $E=0$ to
very high orders in $E$ has been given by Derrida et al. \cite{derrida}.
This analytic expansion however cannot predict singularities in the density
of states.

\begin{figure}
\resizebox{8cm}{!}{
\includegraphics{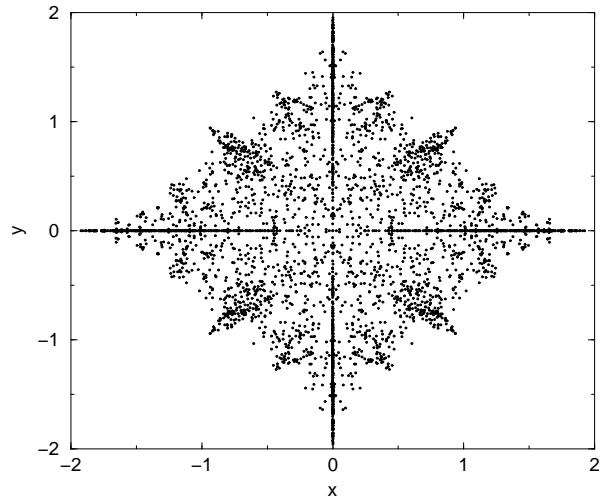}}
\caption{\label{fig2a} Support of the density of states for model A
for a $4000\times4000$ matrix and a single realization of disorder.}
\end{figure}

\begin{figure}
\resizebox{8cm}{!}{
\includegraphics{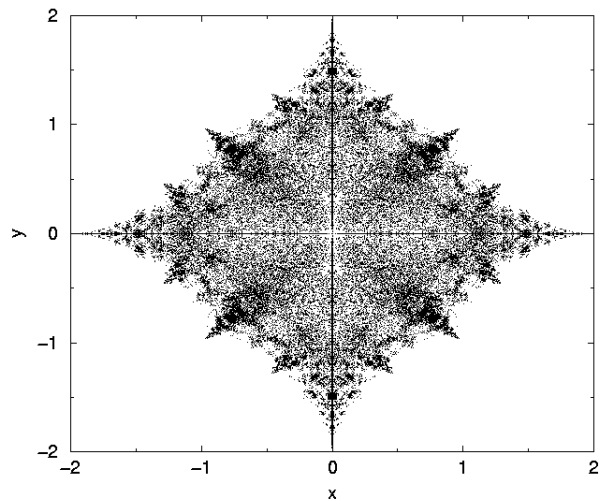}}
\caption{\label{fig2b}Support of the density of states for model A
for a $1000\times1000$ matrix averaged over 100 realizations of disorder.}
\end{figure}

In contrast, for model A the spectrum enjoys only a rectangular $%
Z_{2}\otimes Z_{2}$ symmetry. The first $Z_{2}$ corresponds to $E\rightarrow
E^{\ast }$ obtained by complex conjugating the eigenvalue equation $H\psi
=E\psi .$ The second $Z_{2}$ corresponds to $E\rightarrow -E$ obtained by
the bipartite transformation $\psi _{k}\rightarrow (-)^{k}\psi _{k}.$
Remarkably, FZ found that the spectrum has an enormously complicated
fractal-like form. In Fig.~\ref{fig2a} we plot the support of the density of
eigenvalues in the complex plane for a $4000\times 4000$ matrix, for
a specific realization of the disorder. 
In Fig.~\ref{fig2b} we plot the
support
of the density of states in the complex plane for a $1000\times1000$ matrix,
averaged over 100 realizations of the disorder.

Contrasting Figs.~\ref{fig2a} and~\ref{fig2b} with
Fig.~\ref{fig1}, one can see why it has been a challenge
for mathematical physicists to understand the nature of the
spectrum.

\section{Basic Formalism}

In general, for random non-Hermitian matrices $H$, the density of
eigenvalues can be obtained by
\begin{equation}
\rho (x,y)=\frac{1}{\pi }\frac{\partial }{\partial z^{\ast }}G(z,z^{\ast })
\label{rho}
\end{equation}
where the Green's function is defined by
\[
G(z,z^{\ast })=\left\langle{\frac{1}{N}\mathrm{tr}\frac{1}{z-H}}\right\rangle
\]
with the bracket denoting averaging and $z=x+iy$ (see, for
example, ref.~\cite{8} for a proof of these
relations). Equation~(\ref{rho}) follows from the identity
\begin{equation}
\frac{\partial }{\partial z^{\ast }}\frac{\partial }{\partial z}\mathrm{log}%
\text{ }z=\frac{\partial }{\partial z^{\ast }}\frac{1}{z}=\pi\,\delta
(x)\,\delta (y)\label{Cauchy}
\end{equation}
where $z=x+iy$.
Expanding $G(z,z^{\ast })=\frac{1}{z}\sum_{k=0}^{\infty }{z^{-k}}%
\left\langle{\frac{1}{N}\mathrm{tr}H^{k}}\right\rangle $ we see that $\left\langle{\frac{1}{N}
\mathrm{tr}H^{k}}\right\rangle $ counts the number of paths of a
particle returning to the origin in $k$ steps.

Evidently, for model A each link has to be traversed 4
times. The spectrum of model A was studied by Cicuta et
al. \cite{cicuta} and by Gross and one of us \cite{GZ} by
counting paths. In particular, Cicuta et al. gave an
explicit expression for the number of paths. Recently, the
more general problem of even-visiting random walks has been
studied extensively (see ref. \cite{Bau}). In addition,
exact analytic results for the Lyapunov exponent have been
found by Sire and Krapivsky~\cite{sire}.

In this paper we propose a different approach based on the
theory of words.  We will focus on model A although some of
our results apply to the general class of models described
by (\ref{schrody}).

One important issue is the dimensionality of the spectrum of
model A. In general, the spectrum of non-Hermitian random
matrices when averaged over the randomness is 2-dimensional
(for example, the spectrum of model B). Many of the authors
who have looked at model A believe that its spectrum, as
shown in Figs.~\ref{fig2a} and~\ref{fig2b}, is quasi-0-dimensional: the spectrum
seems to consist of many accumulation points. Here we claim
that the dimension of the spectrum actually lies between $1$
and $2$ dimensions, in the sense described below.

\section{Distribution of the Characteristic Ratio}

Consider the degree $(N+1)$ characteristic polynomial $\Delta
_{N+1}(z)\equiv \mathrm{\det }(H_{N}+zI_{N+1})$, where $I_{N+1}$ 
denotes the $(N+1)\times(N+1)$ identity matrix. We easily
obtain the recursion relation
\[
\Delta _{N+1}(z)=z\Delta _{N}(z)-r_{N}\Delta _{N-1}(z)
\]
with $\Delta _{0}(z)=1$ and $\Delta _{1}(z)=z.$ Note that this second order
recursion relation can be expressed in terms of transfer matrices as

\begin{mathletters}
\begin{equation}
\left(
\begin{array}{c}
\Delta _{N+1}(z) \\
\Delta _{N}(z)
\end{array}
\right) =\left(
\begin{array}{cc}
z & -r_{N} \\
1 & 0
\end{array}
\right) \left(
\begin{array}{c}
\Delta _{N}(z) \\
\Delta _{N-1}(z)
\end{array}
\right)  \label{TM}
\end{equation}
very similar to those defined for the wave function $\psi $,
with the transfer matrix given by
\end{mathletters}
\begin{equation}
U_{N}=\left(
\begin{array}{cc}
z & -r_{N} \\
1 & 0
\end{array}
\right).  \label{TM1}
\end{equation}

Following Dyson and Schmidt~\cite{Dyson,Schmidt} we consider the
characteristic ratio
\begin{equation}
y_{k}(z)\equiv \frac{\Delta _{k}(z)}{\Delta _{k-1}(z)},  \label{y}
\end{equation}
which satisfies the recursion relation
\begin{equation}
y_{k+1}=z-\frac{r_{k}}{y_{k}}  \label{recursion}
\end{equation}
with initial condition $y_1=z$.

The hope is that, while the characteristic polynomials $\Delta _{k}(z)$
obviously changes dramatically as $k$ varies, the characteristic ratio $%
y_{k}(z)$ might converge asymptotically. 

From the definition in eq.~(\ref{y}), it is clear that a point $z$ is in the
spectrum of the $N \times N$ matrix iff $y_N(z)=0$ and
$y_{N+1}(z)=\infty$.
Thus a point $z$ belongs to the spectrum if the
corresponding set of variables $\{y_N\}$ is unbounded, that is, if the
probability of escape of the variable $y_n$ to $\infty$ is finite. As
is shown in \cite{Bau}, this condition
is sufficient to determine the spectrum along the real axis ($z \in
R$), but is insufficient in the complex case.

Let $P_{k}(y_{k})$ be the
probability distribution of $y_{k}$ (note that $y_{k}$ is complex). Then
\[
P_{k+1}(y_{k+1})=\left\langle{\int d^{2}y_{k}\,P_{k}(y_{k})\delta^{(2)}(y_{k+1}-z+
\frac{r_{k}}{y_{k}})}\right\rangle
\]
where the brackets denote the average over the disorder
variables $\{r_{k}\}$. In the thermodynamic limit
$k\rightarrow \infty $, under fairly general conditions
\cite{osedelec} it can be shown that the probability
distribution has a limit $P(y)$, called the invariant
distribution, which is determined by the self-consistent
equation
\begin{eqnarray*}
P(y)&=&\left\langle{\int\,d^{2}t\,P(t)\delta
^{(2)}(y-z+\frac{r}{t})}\right\rangle\\
&=&\left\langle
{\frac{\mid r\mid ^{2}}{\mid z-y\mid ^{4}}P(\frac{r}{z-y})}\right\rangle,
\end{eqnarray*}
where we have used the fact that $\delta ^{(2)}(f(z))$ near
a zero, $z_{0}$,
of $f(z)$ is given by $\delta ^{(2)}(z-z_{0})\,|\frac{df}{dz}|^{-2}$.

For model A, we obtain the amusing equation
\begin{equation}
P(y)=\frac{1}{2\mid z-y\mid ^{4}}\left( P({\frac{1}{z-y}})+P({\frac{-1}{z-y}
})\right).  \label{DS}
\end{equation}

This type of equation has been studied extensively for the
real case in \cite{aeppli,orland,luck}. It can be shown that it can
be solved by the Ansatz
\begin{equation}
P(y)=\sum_{j}a_{j}\delta ^{(2)}(y-b_{j}(z)),  \label{ansatz}
\end{equation}
where the $b_{j}$'s depend on $z$ whereas the $a_{j}$'s don't. Note
that the index $j$ is not necessarily an integer and can refer to
a continuous set. In addition,
it can be shown that the $b_{j}(z)$ are the stable fixed
points of the product of any sequence of transfer matrices
$U$ (see the next section).

Plugging (\ref{ansatz}) into (\ref{DS}) we have
\begin{eqnarray*}
\sum_{j}a_{j}\delta ^{(2)}(y-b_{j})=\frac{1}{2}\sum_{j}a_{j}&&\!\!\!\!\!\!\{\delta
^{(2)}(y-z+\frac{1}{b_{j}})\\
&&+\delta ^{(2)}(y-z-\frac{1}{b_{j}})\}.
\end{eqnarray*}
Since the right hand side has twice as many delta function spikes as the
left hand side, for the two sides to match we expect that, in general, the
index $j$ would have to run over an infinite set.

For a given complex number $z$, we demand that the two sets of complex
numbers $\{b_{j}\}$ and $\{z-1/b_{j},z+1/{b_{j}}\}$ be
the same. This very stringent condition should then determine the $%
\{b_{j}(z)\}.$

To see how this works, focus on a specific $b_{1}$ (Since the label $j$ has
not been specified this can represent any $b_{j}$). It is equal to either $z-%
{1}/{b_{2}}$ or $z+{1}/{b_{2}}$ for some $b_{2}.$ But $b_{2}$ must
in its turn be equal to either $z-{1}/{b_{3}}$ or $z+{1}/{b_{3}}$
for some $b_{3}$. This process of identification must continue until we
return to $b_{1}$. Indeed, if the process of return to $b_{1}$ occurs in a
finite number $L$ of steps, then it will repeat indefinitely (since the
system is back at its starting point $b_{1})$. 
%As was shown in \cite{aeppli}%
%, \cite{orland}, \cite{luck}, 
It is this infinite repetition which gives a
finite weight to the $\delta -$function at $b_{1}$. By contrast, if the
number of steps needed to return to the initial point $b_{1}$ is infinite,
then the weight associated with this point vanishes, and it will not be
present in the spectrum. 
%In fact, it has been shown in the above mentioned
%references, that the weight $a_{j}$ associated to the fixed point $b_{j}$
%corresponding to a repetitive sequence of length $L$ satisfies

We thus conclude that the support of the distribution of $P(y)$ is the
closure of the set of all the stable fixed points of the product of
any sequence of transfer matrices $U$.
We also conclude that the support of the density of states
of the non-Hermitian matrix, in the thermodynamic limit, is given by the zeroes
of any stable fixed point: $b_j(z)=0$.
%
%\begin{equation}
%a_{j}\sim \frac{1}{2^{L}}  \label{a}
%\end{equation}
What is important to notice is that the $a_{j}$ are independent of $z$ and
depend only on the length of the word.
Thus, we conclude that the set of complex numbers $\{b_{j}(z)\}$ is
determined by the solution of an infinite number of fixed point equations.

\section{The Theory of Words}

It is useful here to introduce the theory of words. A word
$w$ of length $L$ is defined as the sequence
$\{w_1,w_2,\cdots ,w_L\}$ where the letters $w_j=\pm 1$.  In
other words, we have a binary alphabet. Let us also define
the repetition of a given word a specific number of times as
a simple sentence. We can then string together simple
sentences to form paragraphs.

For a given word $w$ of length $L$, consider a function $f_{L}(b;z,w)$ to be
constructed iteratively. For notational simplicity we will suppress the
dependence of $f_{L}$ on $b,$ $z,$ and $w,$ indicating only its dependence
on the length $L$ of the word $w.$ The iteration begins with
\[
f_{1}=z-\frac{w_{1}}{b}
\]
and continues with
\begin{equation}
f_{j+1}=z-\frac{w_{j+1}}{f_{j}}.
\end{equation}
We define $f(b;z,w)\equiv f_{L}.$

The set of complex numbers $\{b_{j}(z)\}$ are then determined as follows.
Consider the set of all possible words. For each word $w$, determine the
solution of the fixed point equation
\[
b=f(b;z,w).
\]
By considering small deviations from the solution, we see that the solution
is a stable fixed point only if
\begin{equation}
\left|\frac{\partial f(b;z,w)}{\partial b}\right|<1.  \label{stable}
\end{equation}
The set of all possible words $w$ generates the set of complex numbers $%
\{b_{j}(z)\}.$
In other words, $b$ is determined by a continued fraction equation, since
\[
f(b;z,w)=z-\frac{w_{L}}{z-\frac{w_{L-1}}{z-\frac{w_{L-2}}{\frac{\ddots }{z-%
\frac{w_{1}}{b}}}}}.
\]
We see that $f_{j}$ has the form
\begin{equation}
f_{j}=\frac{\alpha _{j}b+\beta _{j}}{\alpha _{j-1}b+\beta _{j-1}},  \label{fj}
\end{equation}
with the polynomials $\alpha _{j}$ and $\beta _{j}$ determined by the
recursion relations
\begin{equation}
\alpha _{j+1}=z\alpha _{j}-w_{j+1}\alpha _{j-1}  \label{alpha}
\end{equation}
and
\begin{equation}
\beta _{j+1}=z\beta _{j}-w_{j+1}\beta _{j-1},  \label{beta}
\end{equation}
with the initial condition $\alpha _{0}=1,$ $\alpha _{1}=z,$ $\beta _{0}=0,$
and $\beta _{1}=-w_{1}.$ We notice that $\alpha _{j}$ and $\beta _{j}$
satisfy the same recursion relation as that satisfied by $\Delta _{j}$ with
the correspondence $w_{j+1}\leftrightarrow r_{j}.$ Note also that (\ref
{alpha}) and (\ref{beta}) can be packaged as the matrix equation
\begin{equation}
\left(
\begin{array}{ll}
\alpha _{j+1} & \beta _{j+1} \\
\alpha _{j} & \beta _{j}
\end{array}
\right) =\left(
\begin{array}{ll}
z & -w_{j+1} \\
1 & 0
\end{array}
\right) \left(
\begin{array}{ll}
\alpha _{j} & \beta _{j} \\
\alpha _{j-1} & \beta _{j-1}
\end{array}
\right),
\end{equation}
where the transfer matrix $U_{j}$ defined in the previous
section appears. This is closely related to the
transfer matrix formalism discussed earlier. Indeed,
defining $W_{j}\equiv \left(
\begin{array}{ll}
\alpha _{j} & \beta _{j} \\
\alpha _{j-1} & \beta _{j-1}
\end{array}
\right) $ we have the initial condition $W_{1}=\left(
\begin{array}{ll}
z & -w_{1} \\
1 & 0
\end{array}
\right)$. Hence a given word $w$ of length $L$ can also be characterized
by a matrix
\begin{equation}
W=\left(
\begin{array}{ll}
\alpha & \beta \\
\gamma & \delta
\end{array}
\right),
\end{equation}
where for convenience we have written $\alpha =\alpha _{L},$ $%
\beta =\beta _{L},$ $\gamma =\alpha _{L-1}$, and $\delta =\beta _{L-1}.$

For a given word $w,$ the fixed point value $b$ is determined by the
quadratic equation
\begin{equation}
b=\frac{\alpha b+\beta }{\gamma b+\delta },
\end{equation}
which is the fixed point equation of the homographic mapping associated with
the matrix $M$. The geometric interpretation is clear: The matrix $W$ acts
on $2-$component vectors $v$, and we ask for the set of $v$ such that the
ratio $b$ of the first component to the second component is left invariant
by the transformation. In other words, we look for the projective space left invariant
by the transformation $W$: the fixed point value $b$ defines the direction
of the invariant ray.

Hence $b$ is given by
\begin{equation}
b=\frac{P(z)\pm \sqrt{Q(z)}}{2R(z)},  \label{b}
\end{equation}
with $P,Q,R$ polynomials of degree $2L$ in $z$, where $L$ denotes the length
of the word $w$. Explicitly,
\begin{equation}
P_{L}(z)=\alpha _{L}-\beta _{L-1},
\end{equation}
\begin{equation}
Q_{L}(z)=(\alpha _{L}-\beta _{L-1})^{2}+4\alpha _{L-1}\beta _{L},
\end{equation}
and
\begin{equation}
R_{L}(z)=\alpha _{L-1}.
\end{equation}

The stability condition (\ref{stable}) determines which root of (\ref{b}) is
to be chosen.

We will see shortly that $b(z)$ determines the spectrum. Anticipating this,
we see that if we form a compound word by stringing the word $w$ together
twice (for example, the Japanese word ``nurunuru'') then we expect the
contribution to the spectrum to be the same. But given the preceding
discussion, this is obvious, since if a ray is left invariant by $W,$ it is
manifestly left invariant by $W^{2}.$

\section{Density of Eigenvalues}

Once we have determined $b_j(z)$, how do we extract the density of
eigenvalues?
The eigenvalues $\{\lambda _{i}\}$ of the matrix $H$ are given by $\Delta
_{N}(z)=\Pi _{i=1}^{N}(z-\lambda _{i})$. From (\ref{y}) we have $%
\sum_{k=1}^{N}\log y_{k}\simeq \log \Delta
_{N}(z)=\sum_{i=1}^{N}\log (z-\lambda _{i})$, and thus
\begin{equation}
\int \,d^{2}y\,P(y)\log y=\left<\frac{1}{N}\sum_{i=1}^{N}\log 
(z-\lambda _{i})\right>.  \label{bob}
\end{equation}
Using the identity (\ref{Cauchy}) we can differentiate the right hand side
of (\ref{bob}) to obtain the density of eigenvalues in the complex plane
\[
\rho =\frac{1}{\pi }\frac{\partial }{\partial z^{\ast }}\frac{\partial }{%
\partial z}\int\,d^{2}y\,P(y)\log y.
\]
Plugging in our solution
\[
P(y)=\sum_{j}a_{j}\delta ^{(2)}(y-b_{j})
\]
we finally deduce that $\rho =\frac{1}{\pi }\frac{\partial }{\partial
z^{\ast }}\frac{\partial }{\partial z}\sum_{j}a_{j}\log b_{j}=\frac{1%
}{\pi }\frac{\partial }{\partial z^{\ast }}\sum_{j}a_{j}\frac{1}{b_{j}}\frac{%
\partial b_{j}}{\partial z}.$

Since the $a_{j}$ do not depend on $z$, the spectrum is determined
by the zeroes of the fixed point solutions $b_{j}(z).$

We see from (\ref{b}) that the density of eigenvalues is given as a sum over
$j$ of terms like
%(Henri, I am a bit vague here because I don't know what to
%say about $a$ yet)
%
%reply: I wrote that the a's do not depend on z, and thus what you did is OK
\[
\frac{\partial }{\partial z^{\ast }}\left\{\frac{1}{P(z)\pm \sqrt{Q(z)}}[%
P^{\prime }(z)\pm \frac{Q^{\prime }(z)}{2\sqrt{Q(z)}}]-\frac{R^{\prime }(z)}{%
R(z)}\right\}.
\]
Thus the spectrum consists of isolated poles given by the zeroes of $R$ and
$P(z)\pm \sqrt{Q(z)}$, and of the cuts of $\sqrt{Q(z)}$, and is made of
isolated points plus curved line segments connecting the zeroes of $Q(z).$

\begin{figure}
\resizebox{8cm}{!}{
\includegraphics{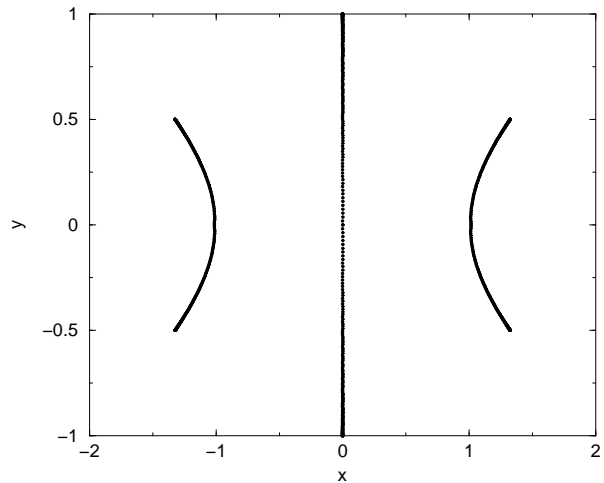}}
\caption{\label{fig3a}Support of the density of states of the periodic
word $\{++-\}$.}
\end{figure}

\begin{figure}
\resizebox{8cm}{!}{
\includegraphics{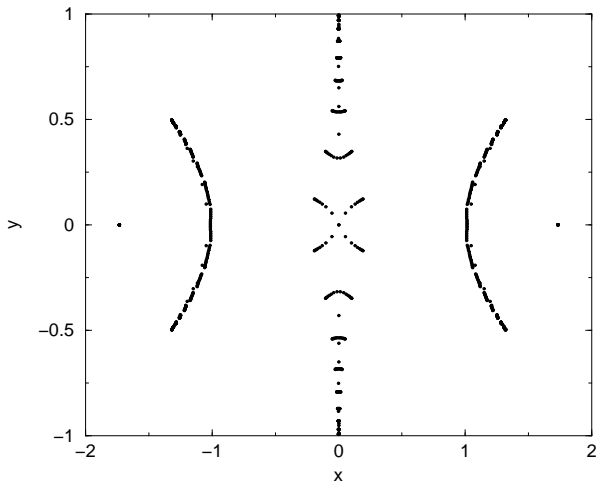}}
\caption{\label{fig3b}Support of the density of states of the periodic word
$\{++-\}$ {\em corrupted}.}
\end{figure}

Contrary to what some authors have believed, the spectrum is not $0-$dimensional,
but $(0+1+\delta)-$dimensional, with $\delta \leq 1$. Each word gives rise
to a line segment, and words which differ slightly from each other gives
rise to line segments near each other. Indeed, given a word $w$, it is
possible to construct a word $w^{\prime }$ with a spectrum as close to that
of $w$ as desired. For that purpose, we may construct $w^{\prime }$ as $
w^{\prime }=w^{l_{1}}vw^{l_{2}}$ where $v$ is any ``corrupting'' word, and
the two lengths $l_{1}$ and $l_{2}$ are sufficiently long. Indeed, in terms of
transfer matrices and invariant rays, we see that $w^{l_{2}}$ acting on any
initial ray brings it close to the stable invariant ray of $w$. Then the
direction of this ray is corrupted by $v$, but it is brought
back arbitrarily close to the invariant ray of $w$ by applying the transfer matrix $
w^{l_{1}}$, provided that $l_{1}$ is large enough. Presumably (although this
remains to be proven rigorously), the spectrum associated with the corrupted
word $w^{\prime }$ can be made as close as we want to that of $w$. We have
thus this property that for any word $w$, there is a word $w^{\prime }$
generating a spectrum as close as we want to that of $w$. In
Figs.~\ref{fig3a} and~\ref{fig3b} we
plot the eigenstates of a word $w=\{++-\}$ and the spectrum of the word $
w^{\prime }=w^{16}\{+++\}w^{16}$. We see that the two spectra are very
close.

As is clear from this discussion, the spectrum is indeed ``incredibly
complicated.''

\section{Words and Spectral Curves}

We content ourselves by focusing on the cuts of $\sqrt{Q(z)}.$ Since for a
word $w$ of length $L,$ $Q(z)$ is a polynomial of degree $2L$ with $2L$
roots, it gives rise to $L$ curved line segments. The curves are given by
the condition
\begin{equation}
\mathrm{Im}\,Q(z)=0.  \label{imq}
\end{equation}
(The sign of \textrm{Re}$\,Q(z)$ depends on the choice for the square root
branch cut.) Given a word $w$ of length $L$, the corresponding
spectrum must be invariant under a cyclic permutation of the letters $%
w_{1},w_{2},\cdots ,w_{L},$ namely under $w_{1}\rightarrow w_{2},$ $%
w_{2}\rightarrow w_{3},\cdots ,w_{L}\rightarrow w_{1}.$

As an example, for the $3-$letter word $w=\{++-\},$ $R(z)=z^{2}+1$ and $%
Q(z)=z^{6}-2z^{4}+z^{2}+4$, which has roots at $z=\pm i,$
$z=\pm (\sqrt{7}-i)/2$, and $z=\pm (\sqrt{7}+i)/2$.

\begin{figure}
\resizebox{8cm}{!}{
\includegraphics{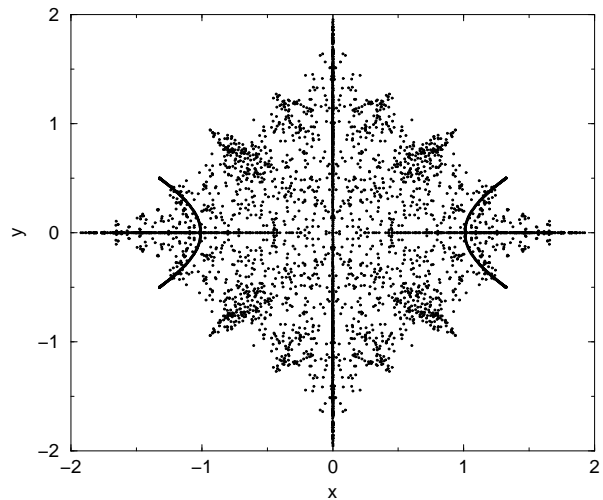}}
\caption{\label{fig4}Support of the density of states of the word $\{++-\}$
over the spectrum of a random $1000\times1000$ matrix}
\end{figure}

It is now clear what the words correspond to ``physically'': a
matrix $H$ with $r_{k}$'s given by an endless repetition of $(+1,+1,-1)$ has
a spectrum given by a straight line connecting $z=\pm i,$ and two algebraic
curves connecting $z=(\sqrt{7}+i)/2$ to $z=(\sqrt{7}-i)/2$
and $z=-(\sqrt{7}+i)/2$ to $z=-(\sqrt{7}-i)/2$, plus poles
at $z=\pm i.$ Notice that the two poles are buried under a cut. In Fig.~\ref{fig4}
we show the spectrum associated with the word $\{++-\}$ together with the
spectrum of a random $1000\times1000$ matrix.

\begin{figure}
\resizebox{8cm}{!}{
\includegraphics{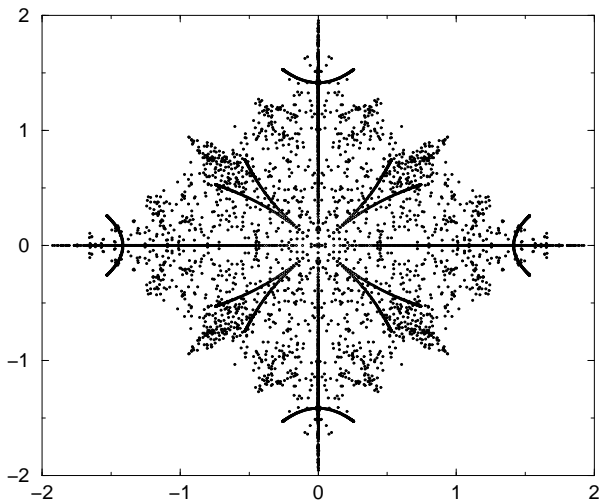}}
\caption{\label{fig5a} Support of the density of states of the periodic
words $\{+++-\}$, $\{++--\}$ and $\{+---\}$
over the support of the spectrum of a random $4000\times4000$
matrix.}
\end{figure}

\smallskip We now give a complete study of all $4-$letter words. The
polynomial is easily found to be
\begin{equation}
Q_{4}(z)=z^{8}-2sz^{6}+(s^{2}+2\kappa )z^{4}-2s\kappa z^{2}+\omega ^{2}
\end{equation}
with $s\equiv w_{1}+w_{3}+w_{2}+w_{4},$ $\kappa \equiv w_{1}w_{3}+w_{2}w_{4}$,
and $\omega \equiv w_{1}w_{3}-w_{2}w_{4}.$ The condition (\ref{imq}) for the
curves in the spectrum reduces to
\begin{equation}
xy\left(y^{2}-x^{2}+\frac{s}{2}\right)\left(x^{4}+y^{4}-6x^{2}y^{2}+s(y^{2}-x^{2})+k\right)=0.
\end{equation}
There are only three non-trivial $4-$letter words, namely $w=\{+++-\},$ $%
\{++--\},$ and $\{+---\}.$ Their contribution 
to the spectrum of $H$ 
together with the spectrum of a random $4000\times4000$ matrix
is shown in Fig.~\ref{fig5a}.

\begin{figure}
\resizebox{8cm}{!}{
\includegraphics{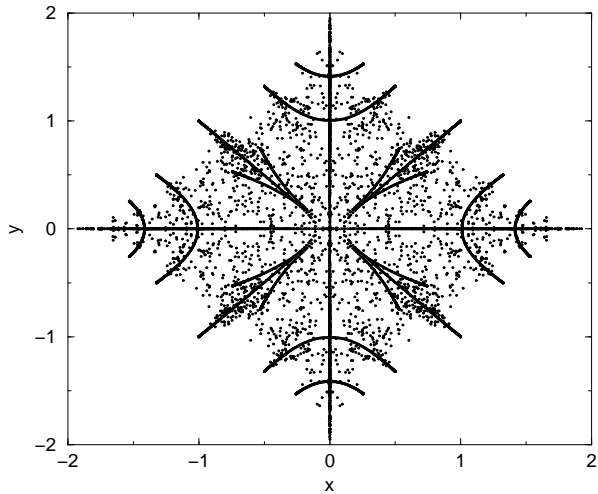}}
\caption{\label{fig5b}Support of the density of states of all periodic
words
of length 4 or less, over the support of the spectrum of a random $4000\times4000$
matrix.}
\end{figure}

In Fig.~\ref{fig5b} we show the contribution of all one, two, three, and four
letter words to the density of states.

\begin{figure}
\resizebox{8cm}{!}{
\includegraphics{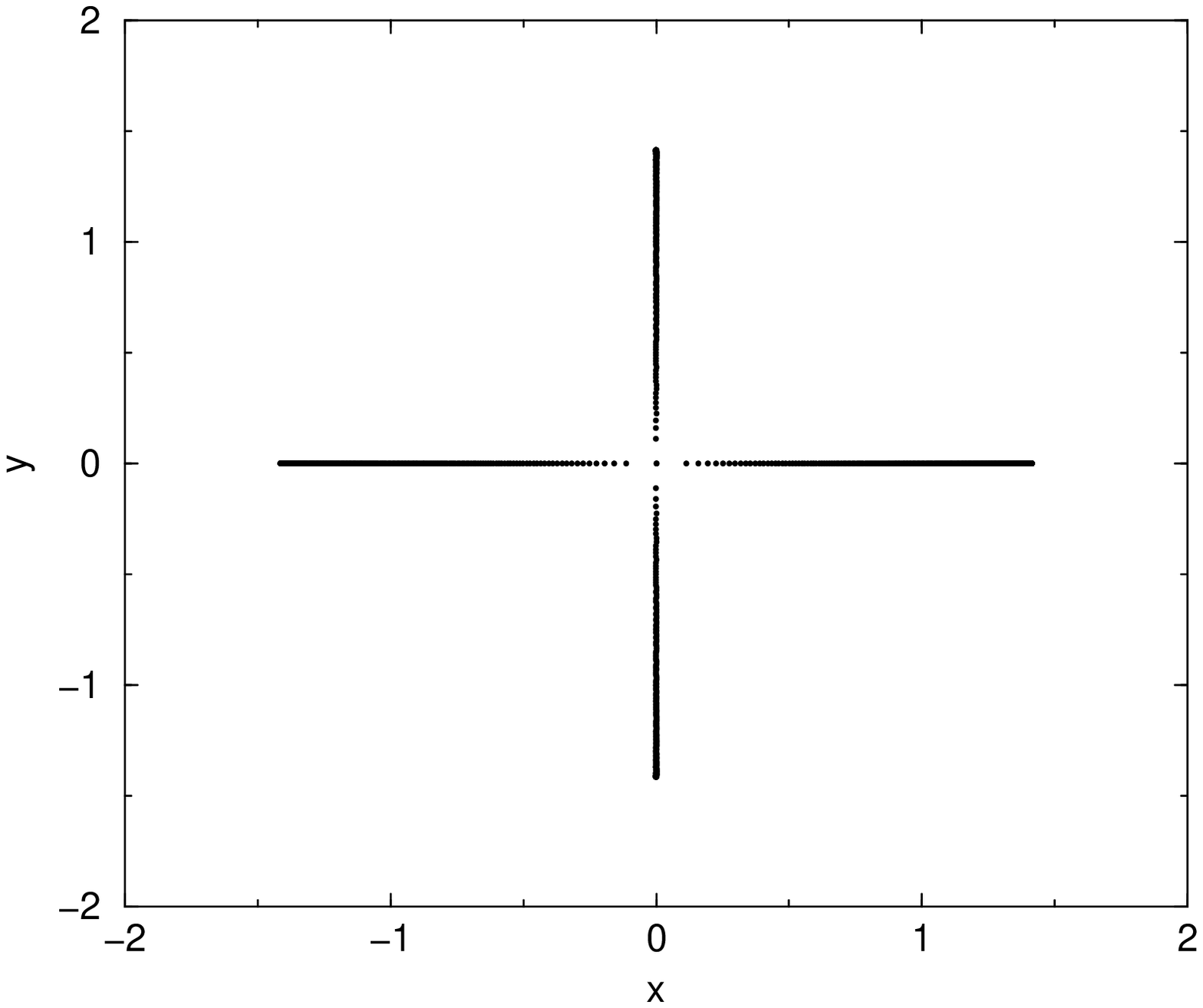}}
\caption{\label{fig6a} Support of the density of states of the periodic word
$\{++--\}$.}
\end{figure}

\begin{figure}
\resizebox{8cm}{!}{
\includegraphics{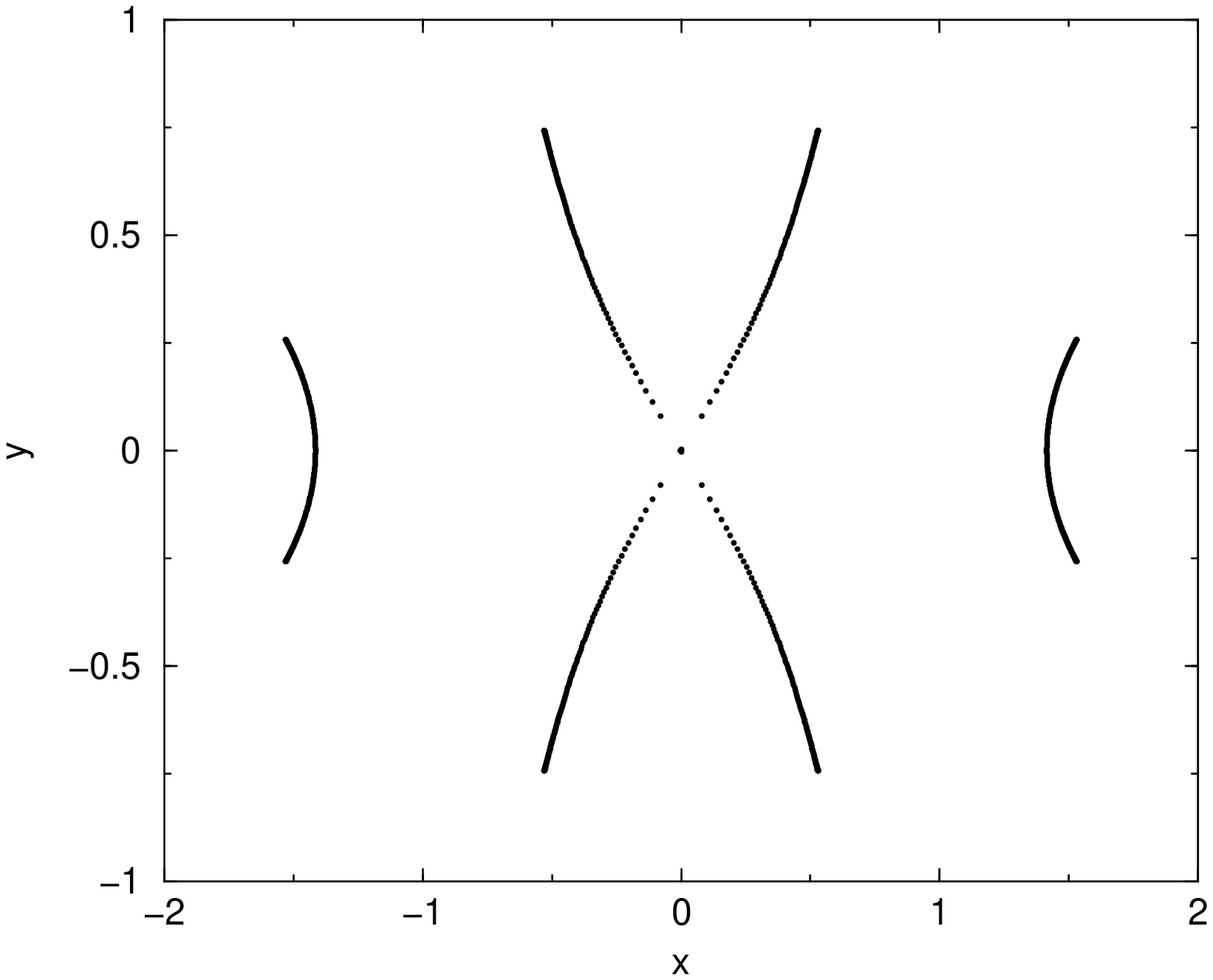}}
\caption{\label{fig6b} Support of the density of states of the periodic word
$\{+++-\}$.}
\end{figure}

\begin{figure}
\resizebox{8cm}{!}{
\includegraphics{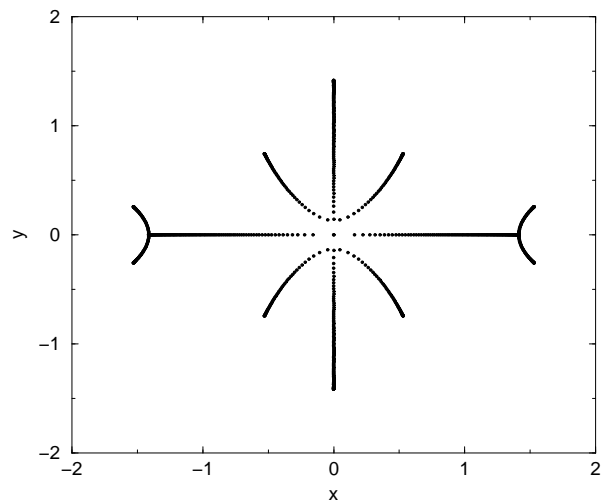}}
\caption{\label{fig6c} Support of the density of states of the
composition of the two words in Figs.~\ref{fig6a} and~\ref{fig6b}.}
\end{figure}

Thus, an $N$ by $N$ matrix $H$ with $r_{k}$'s given by repeating the word $w$
has a spectrum determined by the stable fixed point value $b(z)$
corresponding to $w.$ Furthermore, consider an $N$ \ by $N$ matrix $H$ with $%
r_{k}$'s given by first repeating the word $w_{1}$ (of length $L_{1})$ $%
N_{1} $ times and then by repeating the word $w_{2}$ (of length $L_{2})$ $%
N_{2}$ times. As we would expect, in the limit in which $N_{1}$, $N_{2},$
and $N=N_{1}L_{1}+N_{2}L_{2}$ all tend to infinity, the spectrum of $H$
is given by superposing the spectra of $H_{i}$ $(i=1,2)$, where $H_{i}$ is
constructed with $r_{k}$'s given by first repeating the word $w_{i}$ $N_{i}$
times. This clearly generalizes. In Figs.~\ref{fig6a}--\ref{fig6c} we show the spectrum of the
word $w_{1}=\{++--\}$, the spectrum of $w_{2}=\{+++-\}$, and the spectrum of
the word $w=w_{1}^{20}w_{2}^{20}$. 
We see the superposition principle at work.

From this discussion it becomes clear why the spectrum of the matrix $H$ in
FZ is so complicated. The sequence $\{r_{1},r_{2},\cdots r_{\infty }\}$ is a
book written in the binary alphabet that, in the mathematical limit $%
N\rightarrow \infty $, contains all possible words, sentences, and
paragraphs. In fact, $H$ contains everything ever written or that will be
written in the history of the universe, including this particular paper.
This familiar mind boggling fact accounts for the complicated looking
spectrum first observed in FZ.

It also explains why numerical studies of the spectrum suggest that it is $%
0-$dimensional. Even for $N$ as large as $1000$ the sequence
contains an infinitesimally small subset of the set of all possible words, sentences,
and paragraphs.

\section{Eigenvalues on the Unit Circle}

In the ensemble of all books there are particularly simple books such that $%
\{r_1,r_2,\cdots r_\infty \}$ consists of a word $w$ of length $L$ repeated
again and again. In this case, we can determine the spectrum explicitly by
two different methods.

Let $W$ be the transfer matrix corresponding to $w.$ In other words, $%
W=\prod\limits_{j=1}^{L}\left(
\begin{array}{ll}
z & -w_{j} \\
1 & 0
\end{array}
\right) ,$ where the matrix product is ordered. After repeating the word $R$
times, we have
\begin{equation}
\left(
\begin{array}{l}
\Delta _{RL+1} \\
\Delta _{RL}
\end{array}
\right) =W^{R}\left(
\begin{array}{l}
\Delta _{1} \\
\Delta _{0}
\end{array}
\right).
\end{equation}
Diagonalizing $W=S^{-1}\left(
\begin{array}{ll}
\lambda _{1}(z) & 0 \\
0 & \lambda _{2}(z)
\end{array}
\right) S,$ we see immediately that $\Delta _{RL}$ is a linear function of $%
\lambda _{1}^{R}(z)$ and $\lambda _{2}^{R}(z):$%
\begin{equation}
\Delta _{RL}=\alpha \lambda _{1}^{R}+\beta \lambda _{2}^{R}.  \label{drl}
\end{equation}
We remind the reader that all quantities in (\ref{drl}) are functions
of $z.$

The spectrum of $H$ is determined by the zeroes of $\Delta _{RL}(z)$ as $%
R\rightarrow \infty .$ We note that in this limit the solution of
\begin{equation}
\Delta _{RL}(z)=0  \label{zero}
\end{equation}
does not depend on knowing the detailed form of $\alpha $ and $\beta .$
Indeed, (\ref{zero}) implies
\begin{equation}
\lambda _{1}=\left(-\frac{\beta }{\alpha }\right)^{\frac{1}{R}}\lambda _{2}
\end{equation}
or
\begin{equation}
\lambda _{1}^{2}=\pm \left(-\frac{\beta }{\alpha }\right)^{\frac{1}{R}},
\end{equation}
since $\mathrm{\det }\,W=\lambda _{1}\lambda _{2}=\pm 1.$ In the limit $%
R\rightarrow \infty ,$ $(-{\beta }/{\alpha })^{\frac{1}{R}}$ tends
towards a ($z-$dependent) complex number of modulus unity. Thus, we conclude
that
\begin{equation}
|\lambda (z)|=1  \label{unit},
\end{equation}
namely, that the eigenvalues of $W$ lie on the unit circle. This constraint
suffices to determine the eigenvalues of $H.$ Plugging (\ref{unit}), that is
$\lambda =e^{i\theta },$ into the eigenvalue equation
\begin{equation}
\lambda ^{2}-(\mathrm{tr}\,W)\lambda +\mathrm{\det }\,W=0,
\end{equation}
we obtain $(\mathrm{tr}\,W)=2\cos\theta $ if $\mathrm{\det }\,W=+1$
and $(\mathrm{tr}\,W)=2i\sin \theta $ if $\mathrm{\det }\,W=-1$, which we can
combine into the single equation
\begin{equation}
(\mathrm{tr}\,W)=2(\mathrm{\det }\,W)^{\frac{1}{2}}\cos\theta
\label{result}
\end{equation}
after a trivial phase shift.

%Henri: fix the sine.

\begin{figure}
\resizebox{8cm}{!}{
\includegraphics{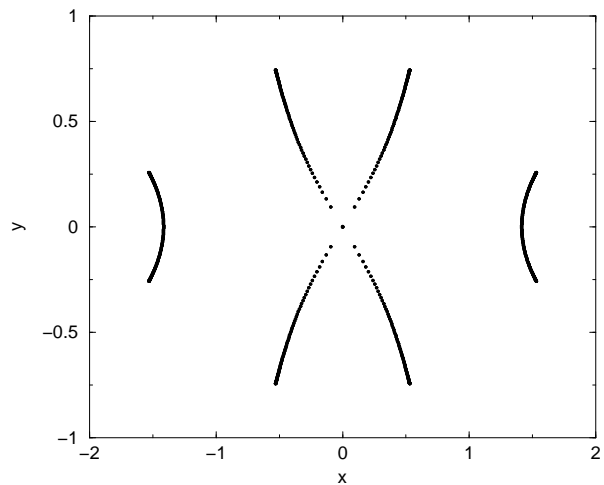}}
\caption{\label{fig7} Support of the density of states of the word $\{+++-\}$
as given by eq. \ref{alg}.}
\end{figure}

As $\theta $ ranges from $0$ to $2\pi ,$
this traces out the spectrum in the complex plane. As an example, consider
the $4-$letter word $\{+++-\}$, in which case (\ref{result}) reduces to
\begin{equation}
\label{alg}
z^{4}-2z^{2}=2i\cos\theta.
\end{equation}
This traces out the algebraic curve shown in
Fig.~\ref{fig7}, which is to be compared to Fig.~\ref{fig6b}.

Of course, since $H$ is now translation invariant with period $L$, we
can apply Bloch's theorem to determine the spectrum of $H.$ Imposing $\psi
_{k+L}=e^{i\varphi }\psi _{k}$ we reduce the eigenvalue problem of $H$ to
the eigenvalue problem of the $L$ by $L$ matrix
\begin{equation}
h_{L}=\left(
\begin{array}{llllll}
0 & 1 & 0 & \cdots & 0 & r_{L}e^{-i\varphi } \\
r_{1} & 0 & 1 & \ddots & 0 & 0 \\
0 & r_{2} & 0 & 1 & \ddots & \vdots \\
\vdots & \ddots & \ddots & \ddots & \ddots & 0 \\
0 & 0 & \ddots & r_{L-2} & 0 & 1 \\
e^{i\varphi } & 0 & \cdots & 0 & r_{L-1} & 0
\end{array}
\right).
\end{equation}
One can verify that with a suitable relation between $\theta $ and $\varphi $
the eigenvalue equation $\mathrm{Det}\,(zI_{L}-h)=0$ becomes identical to (\ref
{result}).

\section{Conclusion and Open Questions}

We have seen that the structure of this simple tridiagonal non-Hermitian
random matrix possesses an amazing richness. This complexity can be understood if
one realizes that the spectrum of the random matrix is the sum of the
spectra of all tridiagonal matrices with a periodic subdiagonal obtained by
repeating an infinite number of times any finite word of
length $L$, weighted by a factor $1/2^{L}$.

The number of lines does not have the cardinal of the
continuum. The number of lines is equal to the number of words of any length that can be made with a
2-letter alphabet; this is a countable number.

There are many open questions concerning the fine structure of the spectrum,
such as whether the spectrum contains holes in the complex plane (in its
domain of definition). We also have not touched upon the question of the
nature of the eigenstates. Are they localized or delocalized? Numerical data
seems to suggest a localization transition. We hope to
address these and other questions in future work.

\section{Acknowledgements}
One of us (HO) would like to thank M. Bauer, D. Bernard and
J.M. Luck for helpful discussions. This work was
partially supported by the NSF under grant PHY99-07949 to
the ITP.

\section{Appendix: Cyclic Invariants}

As remarked in the text, the coefficients of $Q_{L}(z),$ an even polynomial
of degree $2L,$ must be constructed out of the cyclic invariants made of $%
w_{1},w_{2},\cdots ,w_{L}.$ There is presumably a well-developed
mathematical theory of cyclic invariants, but what we need we can easily
deduce here.

For any $L,$ we have the two obvious cyclic invariants $s=\sum_{j=1}^Lw_j$
and $p=\Pi _{j=1}^Lw_j.$ The number of cyclic invariants grows rapidly with $%
L.$ Apparently different cyclic invariants can be constructed out
of other cyclic invariants, for example, $(\sum_{j=1}^Lw_j)^2-(%
\sum_{j=1}^Lw_j^2)=2\sum_{j\neq i}^Lw_jw_i.$

It is easy to work out $Q_{L}(z)$ for low values of $L,$ as follows:
\begin{equation}
Q_{2}(z)=z^{4}-2sz^{2}+d^{2}
\end{equation}
with $d\equiv w_{1}-w_{2}$,
\begin{equation}
Q_{3}(z)=(z^{3}-sz)^{2}-4p=z^{6}-2sz^{4}+s^{2}z^{2}-4p,
\end{equation}
where we have written $Q_{3}(z)$ in a form which shows that its roots can be found
explicitly,
\begin{equation}
Q_{4}(z)=z^{8}-2sz^{6}+(s^{2}+2\kappa )z^{4}-2s\kappa z^{2}+\omega ^{2}
\end{equation}
with $\kappa \equiv w_{1}w_{3}+w_{2}w_{4}$ and $\omega \equiv
w_{1}w_{3}-w_{2}w_{4},$
\begin{equation}
Q_{5}(z)=z^{10}-2sz^{8}+(s^{2}+2\kappa )z^{6}-2s\kappa z^{4}+\kappa
^{2}z^{2}-4p,
\end{equation}
with $\kappa \equiv w_{1}w_{3}+w_{1}w_{4}+w_{2}w_{4}+w_{2}w_{5}+w_{3}w_{5},$
\begin{eqnarray}
Q_{6}(z)&=&z^{12}-2sz^{10}+(s^{2}+2\kappa )z^{8}-2(s\kappa +\rho
)z^{6}\nonumber\\
&&+(\kappa ^{2}+2s\rho )z^{4}-2\kappa \rho z^{2}+\delta ^{2},
\end{eqnarray}
where
\begin{eqnarray}
\kappa
&=&w_{1}w_{3}+w_{1}w_{4}+w_{2}w_{4}+w_{1}w_{5}+w_{2}w_{5}+w_{3}w_{5}\nonumber\\
&&+w_{2}w_{6}+w_{3}w_{6}+w_{4}w_{6}\nonumber\\
\rho &=&w_{1}w_{3}w_{5}+w_{2}w_{4}w_{6}\nonumber\\
\delta &=&w_{1}w_{3}w_{5}-w_{2}w_{4}w_{6}\nonumber
\end{eqnarray}
and
\begin{eqnarray}
Q_{7}(z) &=&z^{14}-2sz^{12}+(s^{2}+2\kappa )z^{10}-2(s\kappa +\rho
)z^{8} \nonumber\\
&&+(\kappa ^{2}+2s\rho )z^{6}-2\kappa \rho z^{4}+\rho ^{2}z^{2}-4p,
\end{eqnarray}
with
\begin{eqnarray}
\kappa &=&w_1w_3+w_1w_4+w_2w_4+w_1w_5+w_2w_5+w_3w_5 \nonumber\\
&&+w_1w_6+w_2w_6+w_3w_6+w_4w_6+w_2w_7+w_3w_7 \nonumber\\
&&+w_4w_7+w_5w_7\nonumber\\
\rho &=&w_1w_3w_5+w_1w_3w_6+w_1w_4w_6+w_2w_4w_6+ \nonumber\\
&&w_2w_4w_7+w_2w_5w_7+w_3w_5w_7\nonumber.
\end{eqnarray}
The quantities $d,\kappa ,\omega ,$ $\kappa ,\rho ,\delta $ are manifestly
cyclic invariants.

%\bigskip


\begin{thebibliography}{99}
\bibitem{FZ}  J.~Feinberg and A.~Zee, Phys. Rev.~E \textbf{59}, 6433 (1999).

\bibitem{1}  N.~Hatano and D.~R.~Nelson, Phys.~Rev.~Lett.~\textbf{77}, 570
(1996).

\bibitem{2}  N.~Hatano and D.~R.~Nelson, Phys.~Rev.~B \textbf{56}, 8651
(1997).

\bibitem{fz2}  J.~Feinberg and A.~Zee, Nucl.~Phys.~B \textbf{552} 599 (1999).

\bibitem{4}  D.~R.~Nelson and N.~M.~Shnerb, Phys.~Rev.~E
\textbf{58}, 1383 (1998).

\bibitem{5}  I.~Y.~Goldsheid and B.~A.~Khoruzhenko, Phys.~Rev.~Lett.~\textbf{%
80}, 2897 (1998).

\bibitem{6}  P.~W.~Brouwer, P.~G.~Silvestrov, and C.~W.~J.~Beenakker,
Phys.~Rev.~B \textbf{56}, R4333 (1997).

\bibitem{7}  K.~B.~Efetov, Phys.~Rev.~Lett.~\textbf{79} 491 (1997).

\bibitem{8}  J.~Feinberg and A.~Zee, Nucl.~Phys.~B \textbf{504} 579 (1997).

\bibitem{9}  E.~Br\'{e}zin and A.~Zee, Nucl.~Phys.~B \textbf{509}, 599
(1998).

\bibitem{10}  A.~Zee, Physica A \textbf{254}, 300 (1998).

\bibitem{11}  N.~Hatano, Physica A \textbf{254}, 317 (1998).

\bibitem{12}  C.~Mudry, P.~W.~Brouwer, B.~I.~Halperin, V.~Gurarie, and
A.~Zee, Phys.~Rev.~B \textbf{58}, 13539 (1998).

\bibitem{derrida}  B.~Derrida, J.~Lykke Jacobsen, and R.~Zeitak,
J.~Stat.~Phys. \textbf{98}, 31 (2000).

\bibitem{cicuta}  G.M. Cicuta, M. Contedini, and L. Molinari, J.~Stat.~Phys.
\textbf{98}, 685 (2000).

\bibitem{GZ}  D. Gross and A. Zee, unpublished.

\bibitem{Bau}  M. Bauer, D. Bernard, and J.M. Luck, J. Phys. A \textbf{34},
2659 (2001).

\bibitem{sire}  C.~Sire and P.L.~Krapivsky, J.~Phys.~A \textbf{34} 9065 (2001).

\bibitem{Dyson}  F.~J.~Dyson, Phys.~Rev.~\textbf{92}, 1331 (1953).

\bibitem{Schmidt}  H.~Schmidt, Phys.~Rev.~\textbf{105}, 425 (1957).

\bibitem{osedelec}  V.I. Oseledec, Trans. Moscow Math. Soc. \textbf{19}, 197
(1968).

\bibitem{aeppli}  R. Bruinsma and G. Aeppli, Phys. Rev. Lett. \textbf{50},
1494 (1983).

\bibitem{orland}  J.M. Normand, M.L. Mehta, and H. Orland, J. Phys. A \textbf{%
18}, 621 (1985).

\bibitem{luck}  J.M. Luck, ``Syst\`{e}mes D\'{e}sordonn\'{e}s
Unidimensionnels (Paris: Collection Al\'{e}a-Saclay, 1992).

\end{thebibliography}
\end{document}